\documentclass[onecolumn, amssymb, preprint, showpacs, nofootinbib, longbibliography]{revtex4-1}    % use J.Chem Phys format

\usepackage{graphicx}			% Include figure files
\usepackage{dcolumn}			% Align table columns on decimal point
\usepackage{bm}					% bold math
\usepackage{float}			% Include float files         https://www.ctan.org/pkg/float?lang=en
\usepackage{amsmath,amsfonts}	% popular packages from the American Mathematical Society
\usepackage{url}					% https://www.ctan.org/pkg/url %-- this is already defined in APS format
\usepackage{setspace}		% Sets spacing    https://www.ctan.org/pkg/setspace
\usepackage{textcomp}
\usepackage{hyperref}
\usepackage{booktabs}
\usepackage{multirow}
\usepackage{subcaption}
\usepackage{epstopdf}
\usepackage[usenames, dvipsnames]{color}
\usepackage{tipa}

\begin{document}
\setcitestyle{authoryear,round,aysep={,}}%Setting citation style. Changing braces to ()
\begin{space}      	% {2.0}

\preprint{The following article has been submitted to Journal of the Acoustical Society of America}		%  if you want want this message to appear in upper left corner of title page

\title[Acoustic Landmarks Contain More Information\ldots]{Acoustic Landmarks Contain More Information About the Phone String than Other Frames for Automatic Speech Recognition with Deep Neural Network Acoustic Model}      % don't need 2 lines, example to show linebreak \\

\author{Di He$^1$}			 \email{dihe2@illinois.edu}
\author{Boon Pang Lim$^2$}			 %\email{boonpang.lim@gmail.com}
\author{Xuesong Yang$^3$}		 %\email{xyang45@illinois.edu}
\author{Mark Hasegawa-Johnson$^3$}		%\email{jhasegaw@illinois.edu}
\author{Deming Chen$^1$}			%\email{dchen@illinois.edu}

\affiliation{$^1$Coordinated Science Lab, University of Illinois at Urbana-Champaign, Urbana, Illinois, USA 61801}

%\author{Author Four}
%\email{author.four@university.edu}
%\thanks{Corresponding author.}
\affiliation{$^2$Novumind, Santa Clara, CA 95054}
 
%\author{Author Five}			% \email{author.five@someplace.edu}
\affiliation{$^3$Beckman Institute, University of Illinois at Urbana-Champaign, Urbana, Illinois, USA 61801}

\date{\today} 
 
\begin{abstract}

Most mainstream Automatic Speech Recognition (ASR) systems consider all feature frames equally important. However, acoustic landmark theory is based on a contradictory idea, that some frames are more important than others. Acoustic landmark theory exploits quantal non-linearities in the articulatory-acoustic and acoustic-perceptual relations to define landmark times at which the speech spectrum abruptly changes or reaches an extremum; frames overlapping landmarks have been demonstrated to be sufficient for speech perception. In this work, we conduct experiments on the TIMIT corpus, with both GMM and DNN based ASR systems and find that frames containing landmarks are more informative for ASR than others. We find that altering the level of emphasis on landmarks by re-weighting acoustic likelihood tends to reduce the phone error rate (PER). Furthermore, by leveraging the landmark as a heuristic, one of our hybrid DNN frame dropping strategies maintained a PER within 0.44\% of optimal when scoring less than half (45.8\% to be precise) of the frames. This hybrid strategy out-performs other non-heuristic-based methods and demonstrate the potential of landmarks for reducing computation.\\
% * <boonpang.lim@gmail.com> 2017-07-25T17:06:01.281Z:
%
% > However, acoustic landmark theory disagrees with this idea.
%
% ^.
% * <boonpang.lim@gmail.com> 2017-07-25T17:05:42.096Z:
%
% >  However, acoustic landmark theory disagrees with this idea.
%
% ^.
% * <boonpang.lim@gmail.com> 2017-07-25T17:04:38.124Z:
%
% > However, acoustic landmark theory disagrees with this idea
%
% ^.

\end{abstract}

%\pacs{PACS: 43.30.Vh, 43.60.Ac, 43.60.Bf, 43.60.Hj }		% From   http://scitation.aip.org/upload/ASA/JASA/JASAAE.pdf  OPTIONAL

\maketitle
%  End of title page -------------------------------------------------------------------------------------------------------------------------------- %
% * <boonpang.lim@gmail.com> 2017-07-25T17:06:08.331Z:
% 
% ABSTRACT -- "However acoustic landmark theory disagrees with this idea" - I think the reviewer complained about this phrasing?
% 
% ^ <dihe2@illinois.edu> 2017-08-03T02:43:44.091Z.

%\linenumbers				% start linenumbers here

\section{Introduction}\label{sec:1} 

 % Recent advances with neural networks have resulted in significant performance gains for modern ASR systems~\citep{xiong2016achieving}, but they may not actually make much use of ideas from speech science. This could be unfortunate as findings from speech science could potentially improve the performance of these systems.
  Ideas from speech science -- which may have the potential to further improve modern
  automatic speech recognition (ASR) -- are not often applied to them~\citep{xiong2016achieving}.
Speech science has demonstrated that perceptual sensitivity to acoustic events is
not uniform in either time or frequency.  Most modern ASR uses a non-uniform
frequency scale based on perceptual models such as 
critical band theory~\citep{fletcher1933loudness}.
In the time domain, however, most ASR systems use a uniform or~\emph{frame synchronous}
time scale: systems extract and analyze feature vectors at regular time intervals, thereby implementing a model according to which the content of every frame is equally important.
%usually with a multi-dimensional log-filterbank spaced evenly along the Mel-frequency domain. %Vectors from adjacent frames may be concatenated  to provide more temporal context. 

%, and rely on a backend decoder to integrate information from frame to frame. 

%Recently, end-to-end systems based on connectionist temporal classification (CTC) for long-short term memory (LSTM) AMs~\citep{sak2015fast} have demonstrated advantages over traditional models such as Gaussian Mixture Models (GMM) and Deep Neural Networks (DNN), moving away from the frame-synchronous approaches which place equal emphasis on each frame. When properly trained with sufficient data, a CTC-LSTM gains the ability to recognize speech, but very little qualitative knowledge can be extracted from the resulting model.

 Acoustic Landmark Theory~\citep{Stevens85,stevens2000acoustic} is a model of 
experimental results from speech science.  It exploits quantal nonlinearities in articulatory-acoustic and acoustic-perceptual relations to define instances in time (landmarks) at which abrupt changes or local extrema occur in speech articulation, in the speech spectrum, or in a speech perceptual response. Landmark theory proposes that humans perceive phonemes in response to acoustic cues, and that such cues are anchored temporally at landmarks, i.e., that a spectrotemporal pattern is perceived as the cue for a distinctive feature only if it occurs with a particular timing relative to a particular type of landmark. Altering distinctive features alters the phone string; distinctive features in turn get signaled by different sets of cues anchored at landmarks. %Assuming Acoustic Landmark information can complement ASR, how can we design experiments to support the hypothesis?
  
The theory of acoustic landmarks has inspired a large number of ASR systems. Acoustic landmarks have been modeled explicitly in ASR system such as those reported in~\citeauthor{hasegawa2005landmark, juneja2004speech, jansen2008hierarchical}.  Many of these systems have accuracies comparable to other contemporaneous systems - in some cases, even returning better performance ~\citet{hasegawa2005landmark}. However, published
landmark-based ASR with accuracy comparable to the state of the art has higher
computation than the state of the art; conversely, landmark-based systems with
lower computational complexity tend to also have accuracy lower than the state of the
art.  No implementation of acoustic landmarks has yet been demonstrated to achieve
accuracy equal to the state of the art at significantly reduced computational 
complexity.  If acoustic landmarks contain more information about the phone string
than other frames, however, then it should be possible to
significantly reduce computational complexity of a state of the art ASR without significantly reducing accuracy, or conversely, to increase accuracy without
increasing computation, by forcing the ASR to extract more information from
frames containing landmarks than from other frames.
%All landmark inspired ASR systems operate on a set of acoustic landmark classifiers that label distinctive features by detecting their correlated acoustic cues. Since acoustic correlates vary dramatically from one distinctive feature to another, there usually does not exist a ``one-for-all'' detector or classifier~\citep{jansen2008hierarchical}. That said, MFCC has demonstrated its effectiveness as a good candidate feature in some applications.~\citet*{qian2016application}, for example, leveraged spliced MFCC features to detect stop consonant landmarks with an error rate lower than 5\%. Results of this sort inspire an attempt to measure the information content of landmarks using an MFCC-based ASR. 
  
We assume that a well trained frame-synchronous statistical acoustic model (AM), having been trained to represent the association between MFCC features and triphones, has also learned sufficient cues and necessary contexts to associate MFCCs and distinctive features. However, because the AM is frame-synchronous, it must integrate information from both informative and uninformative frames, even if the uninformative frames provide no gain in accuracy. The experiments described in this paper explore whether, if we treat frames containing acoustic landmarks as more important than other frames, we can get better accuracy or lower computation.
In this work, we present two methods to quantify the information content of acoustic landmarks in an ASR feature string. In both cases, we use human annotated phone boundaries to label the location of landmarks. The first method seeks to improve ASR accuracy by over-weighting the AM likelihood scores of frames containing phonetic landmarks. By over-weight, we mean multiplying log-likelihoods with a value larger than $1$ (Section~\ref{sec:frame-reweight}). The second method seeks to reduce computation, without sacrificing accuracy, by removing frames from the ASR input. Removing frames makes the computational load decrease, but usually causes accuracy to decrease also; which frames can be removed that cause the accuracy to drop the least? We searched for a strategy that removes as many frames as possible while attempting to keep the Phone Error Rate (PER) low. We show that if we know the locations of acoustic landmarks, and if we retain these frames while dropping others, it is possible to reduce computation for ASR systems with a very small error increment penalty. This method for testing the information content of acoustic landmarks is based on past works~\citep{iso2000speech,vanhoucke2013multiframe,44631} that demonstrated significantly reduced computation by dropping acoustic frames, with small increases in PER  depending on the strategy used to drop frames. In this paper we adopt the PER increment as an indirect measure of the phonetic information content of the dropped frames.
% * <boonpang.lim@gmail.com> 2017-07-25T17:28:53.118Z:
% 
% >  Instead of aiming at a lower PER, we want to compare the PER increment between removing frames with and without landmarks. 
% Furthermore, we take the approach of reducing computation as much as possible with minimal accuracy loss - that is we examine the increase in PER as we remove different types of frames from our decoding.
% 
% ^ <dihe2@illinois.edu> 2017-08-03T03:28:25.064Z.
% * <boonpang.lim@gmail.com> 2017-07-25T17:27:43.190Z:
% 
% > As we will show in later experiments, over-weighting these landmark frames tends to reduce the Phone Error Rate slightly. However, the level of reduction is not significant.
% This one can probably be removed - BP
% Maybe explain what overweighting means exactly (almost to the equation) since it isn't obvious by that.
% 
% ^ <dihe2@illinois.edu> 2017-08-03T03:50:47.377Z.

If the computational complexity of ASR can be reduced without sacrificing accuracy, 
or if the accuracy can be increased without sacrificing computation, these
findings should have practical applications.  It is worth emphasizing that this work only intends to explore these potential applications, assuming landmarks can be accurately detected. Our actual acoustic landmark detection accuracy, despite increasing over time, has not reached a practical level yet.

In this paper, Section~\ref{sec:background} briefly reviews the acoustic landmark theory and relevant works which apply it to ASR systems. Section~\ref{sec:method} presents the theoretical basis for our experiments. Section~\ref{sec:hypo} proposes the hypothesis. Experimental setup is explained in Section~\ref{sec:expe} and results are presented in Section~\ref{sec:results}. Discussion, including a case study of the confusion characteristics is presented in Section~\ref{sec:discussion}. At last, our conclusions are drawn in Section~\ref{sec:conclusion}..

\section{Background and Literature Review}\label{sec:background}

Acoustic landmark theory was first proposed as a theory of the perception
of distinctive features, therefore many landmark-based ASRs use distinctive features
rather than triphones~\citep{lee1988automatic} as their finest-grain categorical representation.
Distinctive features are an approximately binary encoding of perceptual~\citep{Jakobson52}, phonological~\citep{Chomsky68}, and articulatory~\citep{Stevens85} speech sound categories. A feature is called ``distinctive''
if and only if it defines a phoneme category boundary, therefore distinctive features
are language dependent.  The distinctive features used by each language often 
have articulatory, acoustic, and/or perceptual correlates that are similar to those
of distinctive features in other languages, however\citep{Stevens86,stevens2002toward}, 
so it is possible to define a set 
of approximately language-independent distinctive features as follows: if an acoustic
or articulatory feature is used to distinguish phonemes in at least one of the
languages of the world, then that feature may be considered to define a
language-independent distinctive feature.  Each phoneme of a language is 
a unique vector of language-dependent distinctive features.  Automatic speech recognition
may distinguish two different allophones of the same phoneme as distinct
phones; in most cases, the distinctions among phones can be coded using 
distinctive features borrowed from another language, or equivalently, from the 
language-independent set.
  
The ASR community has explored a number of encodings similar to distinctive features, e.g., articulatory features~\citep{Kirchhoff1998,Kirchhoff1999,Kirchhoff2002,Livescu2007,metzearticulatory,Naess2011} and speech attributes~\citep{lee2007overview}. These concepts have different foci, but are also very similar. Distinctive features are defined by phoneme distinctions, 
therefore they are language dependent.  It is possible to define a language-independent
set of distinctive features based on quantal nonlinearities in the 
articulatory-acoustic~\cite{stevens1972quantal} and acoustic-perceptual~\cite{Stevens85}
transformations.  Although both the articulatory-acoustic and acoustic-perceptual
transformations contain quantal nonlinearities that may define distinctive features,
a much larger number of nonlinearities in the articulatory-acoustic transformation
has been demonstrated.  Many studies therefore focus only on the set of phoneme distinctions defined by nonlinearities in the articulatory-acoustic
transformation, which are called ``articulatory features'' in order to denote their
defining principle. Of these, some studies focus on the articulatory-acoustic transform because it implies a degree of acoustic noise robustness~\citep{Kirchhoff1998,Kirchhoff1999,Kirchhoff2002}, others because it implies a compact representation of pronunciation variability~\cite{Livescu2007}, others because it is demonstrably language-independent~\cite{metzearticulatory,Naess2011}. Speech attributes, on the other hand, are a super-set of distinctive features; they are deliberately defined to introduce other purposes to speech recognition. In Lee's framework~\citep{lee2007overview}, speech attributes are quite broadly defined to be perceptible speech categories, of which phonological categories are only a subset. Under this definition, speech attributes include not only distinctive feature but also a wide variety of acoustic cues signaling gender, accent, emotional state and other prosodic, meta-linguistic, and para-linguistic messages.

  As opposed to modern statistical ASR where each frame is treated with equal importance, landmark theory proposes that there exist information rich regions in the speech utterance, and that we should focus on these regions more carefully. These regions of interest are anchored at acoustic landmarks. Landmarks are instantaneous
speech events near which distinctive features are most clearly signaled. These key points mark human perceptual foci and key articulatory events~\citep{liu1996landmark}.~\citet{Stevens85} first introduced these instantaneous
speech events, where, for some phonetic contrasts, humans focus their attention in order to extract acoustic cues necessary for identifying the underlying distinctive features. Initially Stevens named these key points ``acoustic boundaries''; the name ``acoustic landmarks'' was introduced in 1992 (\citeauthor{Stevens1992}), and has been used consistently since. At roughly the same time ~\citet{furui1986role} and~\citet{ohde1994developmental} made similar observations when studying children's speech perception in Japanese. 
  
\citep{liu1996landmark} demonstrated algorithms for automatically detecting
acoustic landmarks.  \citep{hasegawa2000time} measured the phonetic information
content of known acoustic landmarks.
In his work, \citet{hasegawa2000time} defined a set of landmarks including consonant releases and closures (at phone boundaries) and vowel/glide pivot landmarks (near the center of the corresponding phones). In contrast,~\citet{lulich2010subglottal} argued that the center of vowels and glides are not as informative and should not be considered as landmarks. He defined, instead, formant-subglottal resonance crossing, which is known to sit between boundaries of [-back] and [+back] vowels, to be more informative. In a paper, ~\citet{wang2009automatic} showed that the latter improves performance for automatic speaker normalization application. 
In~\citep{hasegawa2000time},
a small number of pivot and release landmarks were defined at $+33\%$ and $-20\%$ locations after the beginning or before the end of certain phones (where $+33\%$ indicates delaying the location by $33\%$ of the total duration of that phone; $-20\%$ indicates advancing the location by $20\%$), in order to better approximate the typical timing of the spectrotemporal events defined earlier in Liu's work~\citep{liu1996landmark}. Later works~\citep{hasegawa2005landmark, xiang2016landmark} labeled these landmarks right on the boundary and returned similar performance with~\citet{hasegawa2000time}.  Figure \ref{fig:landmark_eg} illustrates the landmark labels for the pronunciation of word ``Symposium''\footnote{The pronunciation of ``Symposium'' is selected from audio file: \url{TIMIT/TRAIN/DR1/FSMA0/SX361.WAV}}. The details landmark labeling heuristics applied in this example are further described in Table~\ref{tab:landmark_rules}.

\begin{figure}[htbp]
	\centering
    \includegraphics[width=\textwidth]{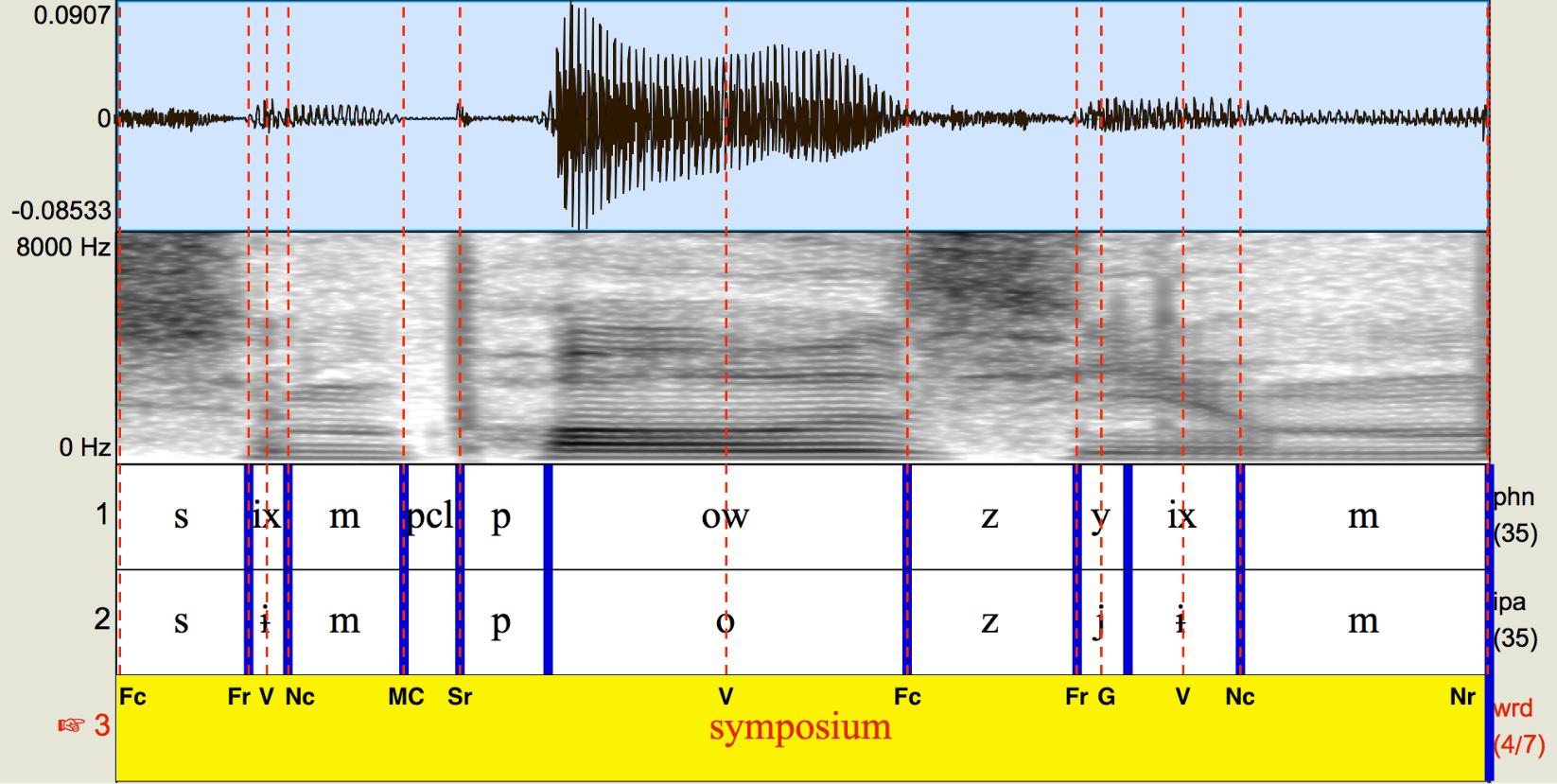}
    \caption{Acoustic landmark labels (LM) for the pronunciation of the word ``Symposium''. TIMIT phone symbols (PHN) and international phonetic alphabet (IPA) symbols are both used in this example. The dashed red lines denote the landmark positions. The symbols under the dashed red lines are landmark types, where \textbf{Fc} and \textbf{Fr} are closure and release for fricatives; \textbf{Sc} and \textbf{Sr} are closure and release for Stops; \textbf{Nc} and \textbf{Nr} are closure and release for nasals; \textbf{V} and \textbf{G} are vowel pivot and glide pivot; \textbf{MC} is manner-change landmark.}
    \label{fig:landmark_eg}
\end{figure}

 Many works have focused on accurately detecting acoustic landmarks. The first of these assumed that landmarks correspond to the temporal extrema of energy or energy change in particular frequency bands, e.g.,~\citet{liu1996landmark} detected consonantal landmarks in this way,~\citet{Howitt2000} detected vowel landmarks,~\citet{Choi1999} classified consonant voicing, and~\citet{LeeJ2008,LeeS2008,Lee2011,Lee2012} classified place of articulation. Support Vector Machines (SVMs) were popularized for landmark detection by~\citet*{Niyogi99}, who showed that an SVM trained to observe a very small acoustic feature vector (only four measurements, computed once per millisecond) can detect stop release landmarks more accurately than a hidden Markov model. Both~\citet{borys2008svm} and~\citet{chitturi2006novel} target the detection of all landmarks using one kind of acoustic features. Their results are reasonably accurate, but are still less accurate and more computationally expensive than the best available classifier for each distinctive feature.~\citet*{xie2006robust} expanded the work of~\citet*{Niyogi99} by demonstrating detection of several different types of landmark using a very small acoustic feature vector. In Qian's paper~\citep{qian2016application}, a small vector of acoustic features was learned, using the technique of local binary patterns, and resulting in accuracy above 95\% for stop consonant detection. In a paper from~\citet{xiang2016landmark}, a Convolutional Neural Network (CNN) trained on MFCC and additional acoustic features achieved around 85\% on consonant voicing detection. This system was trained on the English corpus TIMIT~\citep{garofalo1993darpa}, but tested on Spanish and Turkish corpora. Over time, new techniques and more specific features have been developed for landmark detection, and the detection accuracy has been improving steadily. 
Acoustic landmarks were first introduced as part of an ASR in 1992 (\citeauthor{Stevens1992}), and have been used in a variety of ASR system architectures.  These systems, without considering the mechanism used for landmark detection, can be clustered into two types. The first type of system, such as those described by~\citet{liu1996landmark,jansen2008hierarchical,juneja2004speech} computes a lexical transcription directly from a set of detected distinctive features. Due to the complexity of building a full decoding mechanism on distinctive features, some of these systems only output isolated words. However, other systems (e.g.,work from~\citet{jansen2008hierarchical}) have full HMM back-ends that can output word sequences. The other type of system, such as that described by~\citet{hasegawa2005landmark}, conducts landmark-based re-scoring on the lattices generated by an MFCC-based hidden Markov model. Acoustic likelihoods from the classic ASR systems are adjusted by the output of the distinctive feature classifier. Many landmark based ASRs demonstrated performance slightly~\citep{hasegawa2005landmark} or even significantly~\citep{Kirchhoff1998} better than baseline ASR systems, especially in noisy conditions.

\section{Measures of the Information Content of Acoustic Frames}\label{sec:method}

An acoustic landmark is an instantaneous event that serves as a reference time point for the measurement of spectrotemporal cues widely separated in time and frequency. For example, in the paper that first defined landmarks, Stevens proposed classifying distinctive features of the landmark based on the onsets and offsets of formants and other spectrotemporal cues up to 50ms before or 150ms after the landmark~\citep{Stevens85}. The 200ms spectrotemporal dynamic context proposed by Stevens is comparable to the 165ms spectrotemporal dynamic context computed for every frame by the ASR system of~\citet{Vesely13interspeech}. Most ASR systems use acoustic features that are 
derived from frames 25ms long, with a 10ms skip, as human speech is quasi-stationary for this short period~\citep{quatieri2008discrete}. Because spectral dynamics communicate distinctive features, however, ASR systems since 1981 (\citeauthor{Furui81}) have used dynamic features; since deep neural nets (DNNs) began gaining popularity, the complexity of the dynamic feature set in each frame has increased quite a lot, with consequent improvements in ASR accuracy.   This trend not only applies to stacking below 100ms. With careful normalization, features like TRAPs~\citep{hermansky2003trap}, with temporal window equal or longer than 500ms, continue to demonstrate accuracy improvement.  Experiments reported in this paper are built on a baseline described by~\citet{Vesely13interspeech}, and schematized in Fig.~\ref{fig:stack}.  In this system, MFCCs are computed once every 10ms, with 25ms windows (dark gray rectangles in Fig~\ref{fig:stack}). In order to include more temporal context, we stack adjacent frames, three preceding and three succeeding, for a total of seven frames (a total temporal span…of $(7-1)\times 10+25=85$ ms). These are shown in Fig~\ref{fig:stack} as the longer, segmented dark gray rectangles, with the red segments representing the center frames of each stack.  The seven-frame stack is projected down to 40 dimensions using linear discriminant analysis (LDA). For input to the DNN but not the GMM, LDA is followed by speaker adaptation using mean subtraction and feature-space maximum likelihood linear regression, additional context is provided by a second stacking operation afterwards, in which LDA-transformed features, represented by yellow rectangles, are included in stacks of 9 frames (for a total temporal span of $(9-1)\times 10+85=165$ms), as represented by the top path in Fig~\ref{fig:stack}. It is believed that the reason features spanning longer duration improve ASR accuracy is that long lasting features capture coarticulation better, including both neighboring-phone transitions and longer-term coarticulation. The dynamics of the tongue naturally cause the articulation of one phoneme to be reflected in the transitions into and out of neighboring phonemes, over a time span of perhaps 70ms.  Longer-term coarticulation, spanning one or more syllables, can occur when an intervening phoneme does not require any particular placement of one or more articulators, e.g.,~\citet{Ohman65a} demonstrated that the tongue body may transition smoothly from one vowel to the next without apparently being constrained by the presence of several intervening consonants.
% * <boonpang.lim@gmail.com> 2017-08-05T04:45:15.504Z:
% 
% > arced
% anchored?
% 
% ^ <dihe2@illinois.edu> 2017-08-05T20:03:09.647Z.
% * <boonpang.lim@gmail.com> 2017-07-26T21:36:59.709Z:
% 
% > TRAPs~\citep{},
% citation needed
% 
% 
% ^ <dihe2@illinois.edu> 2017-08-03T03:22:18.495Z.
% * <boonpang.lim@gmail.com> 2017-07-26T21:34:13.105Z:
% 
% > The following 2 paragraph attempts to bridge to the connection of acoustic cue and features extraction from ASR systems. Specifically focusing on explaining how ASR feature extraction incorporates the information necessary for distinctive feature analysis. 
% 
% I don't think you quite need to be so explicit when trying to bridge the sections. Can we do it more subtly?
% 
% ^ <dihe2@illinois.edu> 2017-08-03T06:04:49.490Z:
% 
% Revised the order of paragraph, please offer feedbacks.
%
% ^ <dihe2@illinois.edu> 2017-08-05T20:02:12.816Z.

    \begin{figure}[htbp]
      %\vspace{-2mm}
      \centering
      \includegraphics[width=\textwidth]{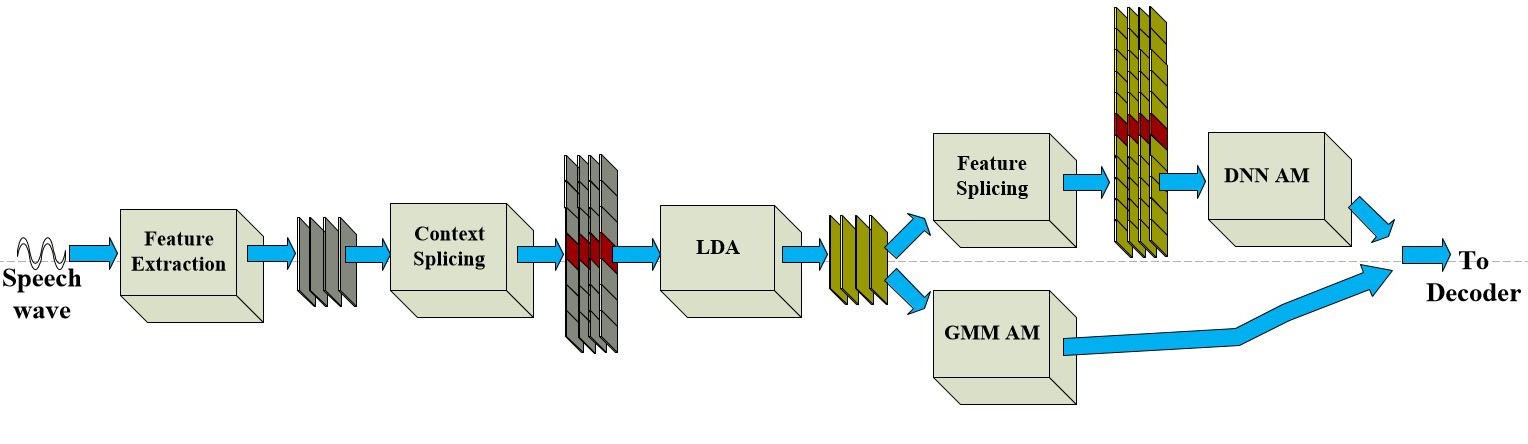}
      \caption{{\it Stacking of Feature Frames Before the Scoring Process for DNN AM (top path) and GMM AM (bottom path). The dark gray, red and green rectangles indicate frames and stacks of frames. Please see text.} \label{fig:stack}}
      %\vspace{-2mm}
    \end{figure}

\subsection{Frame Re-weighting}\label{sec:frame-reweight}

  HMM-based ASR searches the space of all possible state sequences for the most likely state sequence given the observations. During the state likelihood estimation, results of all frames are weighted equally. Weighting more informative frames more heavily could potentially benefit speech recognition. Ignoring the effects of the language model, the log-likelihood of a state sequence $S$ given the observations $O$ is 
\begin{equation} \label{eq:1}
  L(S|O) = \sum^T_{t=1} w(t)log(p(o_t|s_t)) + log(p(s_t| s_{t-1})) ,
\end{equation}
where $s_t$ and $o_t$ are respectively the state and observed feature vector associated with the frame at time~$t$. The state $s_t$ at any time should be associated with one of the senones (i.e., monophone or clustered triphone states). Here $p(s_t| s_{t-1})$ is the transition probability between senones, which we will not consider modifying in this study. In most systems, beam search parameters constrain the number of active states, thus we only need to evaluate the necessary posteriors.  In our over-weighting framework, if $o_t$ contains a landmark, the value of $\log p(o_t | s_t)$ will be scaled. To simplify the computation, we operate directly on log-likelihoods. In this case, $log(p( o_t | s_t))$ is multiplied by factor $w(t)$ which takes the value $1$ when frame $t$ contains no landmark and a value greater than $1$ otherwise. This is effectively applying a power operation on the likelihoods.

  The key in this strategy is that the likelihood of all model states will be re-weighted. If the frame over-weighted is a frame that can differentiate the correct state better, the error rate will drop. In contrast, if the likelihood of a frame is divided evenly across states, or even worse, is higher for the incorrect state, then over-weighting this frame will mislead the decoder and increase chances of error. For this reason, over-weighting landmark frames is a good measure to tell how meaningful landmark frames are compared to the rest of the frames. If the landmarks are indeed more significant, we should observe a reduction in the PER for the system over-weighting the landmark.

\subsection{Frame dropping}
  
  The wide temporal windows used in modern ASR, as mentioned in the beginning of Section~\ref{sec:method}, are highly useful to landmark-based speech recognition:
  all of the dynamic spectral cues proposed by Stevens~\citeyear{Stevens85}
  are within the temporal window spanned by the feature vector of a frame centered 
  at the landmark, therefore it may be possible to correctly identify the distinctive
  features of the landmark by dropping all other frames, and keeping only the frame
  centered at the landmark.
  Our different frame dropping heuristics modify the log probability of a state sequence by replacing the likelihood $p( o_t | s_t)$ with an approximation function $f$. In terms of log probabilities, Equation~\eqref{eq:1} becomes
\begin{equation}\label{eq:2}
L(S|O) = \sum^T_{t=1} log{f(p(o_t|s_t), t)} + log(p(s_t, s_{t-1})),
\end{equation}
The class of optimizations considered in this paper involve a set of functions $f(p(o_t|s_t))$ parameterized as:
\begin{equation}
f(p(o_t|s_t)) = \begin{cases}
     R(O,t)   & \text{if } g(t) = 1 \\
     p(o_t|s_t)        & \text{otherwise}
    \end{cases},
\end{equation}
The \emph{method of replacement} is characterized by $R$, and the frame-dropping function by $g(t)$. This work considers multiple methods to verify that the finding with respect to landmarks is independent of the replacement method. The four possible settings of the $R(o,t)$ function are as follows:
% Xuesong: equations description
\begin{equation}
R(O,t)\in\left\{\begin{array}{ll}
R_{\textrm{Copy}}(O,t) &= p(o_{t'}|s_{t'}),\quad t'=\max_{\tau\le t,~g(\tau)=0}\tau\\
R_{\textrm{Fill\_0}}(O,t) &= 1\\
R_{\textrm{Fill\_const}}(O,t) &= \left(\prod_{t=1}^T p(o_t|s_t)\right)^{1/T}\\
R_{\textrm{Upsample}}(O,t) &= \exp\left(
\sum_{t':g(t')=0} h(t-t')\log p(o_t|s_t)\right)
\end{array}\right.
\end{equation}
In other words, the {\tt Copy} strategy copies the most recent observed value of $p(o_t|s_t)$, the {\tt Fill\_0} strategy replaces the log probability by 0, the {\tt Fill\_const} strategy replaces the log probability by its mean value, and the {\tt Upsample} strategy replaces it by an interpolated value computed by interpolating (using interpolation filter $h(t)$) the log probabilities that have been selected for retention.  The {\tt Upsample} strategy will only be used if the frame-dropping function is periodic, i.e., if frames are downsampled by a uniform downsampling rate.

The \emph{pattern of dropped frames} can be captured by the indicator function $g$, which is true for frames that we want to drop. Experiments will test two landmark-based patterns: {\tt Landmark-drop} drops all landmark frames ($g(t)=1$ if the frame contains a landmark), and {\tt Landmark-keep} keeps all landmark frames ($g(t)=1$ only if the frame does {\em not} contain a landmark). In the case where landmark information is not available, the frame-dropping pattern may be {\tt Regular}, in which $g(t) = \delta( \text{ $t$  mod $ K$ } )$ indicating that every $K$-th frame is to be dropped, or it may be {\tt Random}, in which case the indicator function is effectively a binary random variable set at a desired frame dropping rate. As we will demonstrate later, to achieve a specific function and dropping ratio, we can sometimes combine output of different $g$ functions together by taking a logical inclusive OR to their output. 

  If acoustic landmark frames contain more valuable information than other frames, it can be expected that experiment setups that retain the landmark frames should out-perform other patterns, while those that drop the landmark frames should under-perform, regardless of the \emph{method of replacement} chosen.
  
\section{Hypotheses}\label{sec:hypo}
  
 This paper tests two hypotheses. The first is that a window of speech frames (in this case 9 frames)  centered at a phonetic landmark has more information than windows centered elsewhere -- this implies that over-weighting the landmark-centered windows can result in a reduction in PER\@. The second hypothesis states that keeping landmark-centered windows rather than other windows causes little PER increment, and that dropping a landmark-centered window causes greater PER increment as opposed to dropping other frames. In the study we focused on PER as opposed to Word Error Rate (WER) for two reasons. First, the baseline Kaldi recipe for TIMIT reports PER. Second, this study is oriented towards speech acoustics; focusing on phones allow us to categorize and discuss the experiment and results in better context.
  
  %\begin{table*}[htbp]
  \begin{table}[htbp]
    %\vspace{-2mm}
  	\caption{\label{tab:landmark_rules} {\it Landmark types and their positions for acoustic segments. %where `c', and `r' denote consonant closure, and release; `start', `middle', and `end' denote three positions across acoustic segments, respectively.
\textbf{Fc} and \textbf{Fr} are closure and release for fricatives; \textbf{Sc} and \textbf{Sr} are closure and release for Stops; \textbf{Nc} and \textbf{Nr} are closure and release for nasals; \textbf{V} and \textbf{G} are vowel pivot and glide pivot; `start', `middle', and `end' denote three positions across acoustic segments.}}
  	%\vspace{2mm}
  	\centering
  	\begin{tabular}{||p{2.2cm}|p{4cm}|p{9cm}||}
  		\hline
  		Manner of Articulation & Landmark Type and Position & Observation in Spectrogram\\
  		\hline
  		Vowel & V: middle & maximum in low- and mid-frequency amplitude\\
  		Glide & G: middle & minimum in low- and mid-frequency amplitude\\
        \hline
  		Fricative & Fc: start, Fr: end & \multirow{4}{9cm}{amplitude discontinuity occurs when consonantal constriction is formed or released} \\
  		Affricate & Sr,Fc: start, Fr: end & \\ 
  		Nasal & Nc: start, Nr: end &\\
  		Stop  & Sc: start, Sr: end &\\
  		\hline
  	\end{tabular}
    %\vspace{-4mm}
  \end{table}
  
In order to test these hypotheses, a phone boundary list from the TIMIT speech corpus~\citep{garofalo1993darpa} was obtained, and the landmarks were labeled based on the phone boundary information. Table~\ref{tab:landmark_rules} briefly illustrates the types of landmarks and their positions, as defined by the TIMIT phone segments. This marking procedure is shared by~\citet{stevens2002toward,hasegawa2005landmark,xiang2016landmark}. It is worth mentioning that this definition disagrees with that of~\citet{lulich2010subglottal}. Lulich claims that there is no landmark in the center of Vowel and Glide; instead, a formant-subglottal resonance crossing, which is known to sit between the boundaries of [-Back] and [+Back] vowels, contains a landmark. Frames marked as landmark are of interest. To test hypothesis 1, landmark frames are over-weighted.  To test hypothesis 2, either non-landmark or landmark frames are dropped.
  
  \section{Experimental Methods}\label{sec:expe}

  Our experiments are performed on the TIMIT corpus. Baseline systems use standard examples distributed with the Kaldi open source ASR toolkit\footnote{\url{http://kaldi-asr.org/}}. Specifically, the GMM-based baseline follows the configurations in the distributed \verb$tri2$ configuration in the Kaldi TIMIT example files\footnote{\url{hhttps://github.com/kaldi-asr/kaldi/tree/master/egs/timit/s5}}. The clustered triphone models are trained using maximum likelihood estimation of features
  that have been transformed using linear discriminant analysis and maximum likelihood linear transformation. For the DNN baseline, speaker adaptation is performed on the features, and nine consecutive frames centered at the current frame are stacked as inputs to the DNN, as specified in the distributed ~\verb$tri4_nnet$ example. Respectively, the two systems achieved PER of $23.8\%$ (GMM) and $22.6\%$ (DNN) without any modification.
% * <boonpang.lim@gmail.com> 2017-08-05T04:49:43.722Z:
% 
% > In the study we focused on PER as oppose to Word Error Rate (WER) for 2 reasons, first the baseline Kaldi recipe for TIMIT reports PER; second, this study orients acoustic science, focusing on phones allow us to categorize and discuss the experiment and results in better context.
% This motivation is repeated in the front already, can we move that that so that this describes more of the methodology rather than the motivations.
% 
% ^ <dihe2@illinois.edu> 2017-08-05T20:16:02.218Z.

We performed a 10-fold cross validation (CV) over the full corpus, by first combining the training and test sets, and creating 10 disparate partitions for each test condition. The gender balance was preserved to be identical to the canonical test set for each test subset, while the phonetic balance was approximately the same but not necessarily identical. This is in order to improve the significance of our PER numbers. The TIMIT corpus is fairly small and the phone occurrence of some phones, or even phone categories, in the test set is lower than ideal. Conducting cross validation on the full set allows us partially address this issue.
% * <boonpang.lim@gmail.com> 2017-07-26T21:38:55.892Z:
% 
% > Apart from the default test set, we also
% We performed a 10-fold ...
% 
% ^ <dihe2@illinois.edu> 2017-08-03T03:23:08.147Z.

  For the control experiments of our tests, all configurations of feature extraction and decoding process are retained the same as the baseline. In this case, fair comparisons are guaranteed, and we can fully reveal the effects of our methods in the AM scoring process. 
  
  \section{Experimental Results}\label{sec:results}
  
  Experimental results examining the two hypotheses proposed above will be presented in this section. We will present the results of over-weighting the landmark frames first. Evaluation of frame dropping will be presented second, and includes several phases. In the first phase, a comparison of different \emph{methods of replacement} is presented, to provide the reader with more insight into these methods before they are applied to acoustic landmarks. In the second phase, we will then leverage our findings to build a strategy that both drops non-landmark frames, and over-weights landmark frames, using the best available \emph{pattern of dropped frames} and \emph{method of replacement}.
  
% * <boonpang.lim@gmail.com> 2017-08-05T04:51:58.426Z:
% 
% >  comparison of different \emph{methods of replacement} is presented, to provide the reader with more insight into these methods before they are applied to acoustic landmarks. In the second phase, we will then leverage our findings to build a strategy that both drops non-landmark frames, and over-weights landmark frames, using the best available \emph{pattern of dropped frames} and \emph{method of replacement}.
% >   
% emphasis on the key concepts here should be done when we first introduce them .. i.e. in the sections on the equations.
% 
% ^ <dihe2@illinois.edu> 2017-08-05T20:22:43.217Z.
  \subsection{Hypothesis 1: Over-weighting Landmark Frames}
  \label{sec:over}
  
  Figure~\ref{fig:reweight} illustrates the PER of the strategy of over-weighting the landmark frames during the decoding procedure, and how it varies with the factor used to weight the AM likelihood of frames centered at a landmark. The PER for GMM-based models drops as the weighting factor increases until the factor is 1.5; increasing the weighting factor above 1.5 causes the PER to increase slightly.  When the factor is increased to greater than 2.5, the PER increases at a higher slope. Similar trends can be found for DNN models, yet in this case the change in PER is non-concave and spans a smaller range. If landmark frames are under-weighted, or over-weighted by a factor of 1.5 or up to 2.0, PER increases. Over-weighting landmark frames by a factor of 3.0 to 4.0 reduces PER\@. %Hypothesis tests~\citep{gillick1989some} have been conducted and neither of the PER reductions (GMM or DNN) is statistically significant. 
In this experiment, Wilcoxon tests~\citep{gillick1989some} have been conducted, through Speech Recognition Scoring Toolkit (SCTK) 2.4.10\footnote{\url{https://www.nist.gov/itl/iad/mig/tools}}, and tests concluded the difference to be insignificant.
  
    %\vspace{-0.5mm}
    \begin{figure}[htbp]
      \centering
      \includegraphics[width=\linewidth]{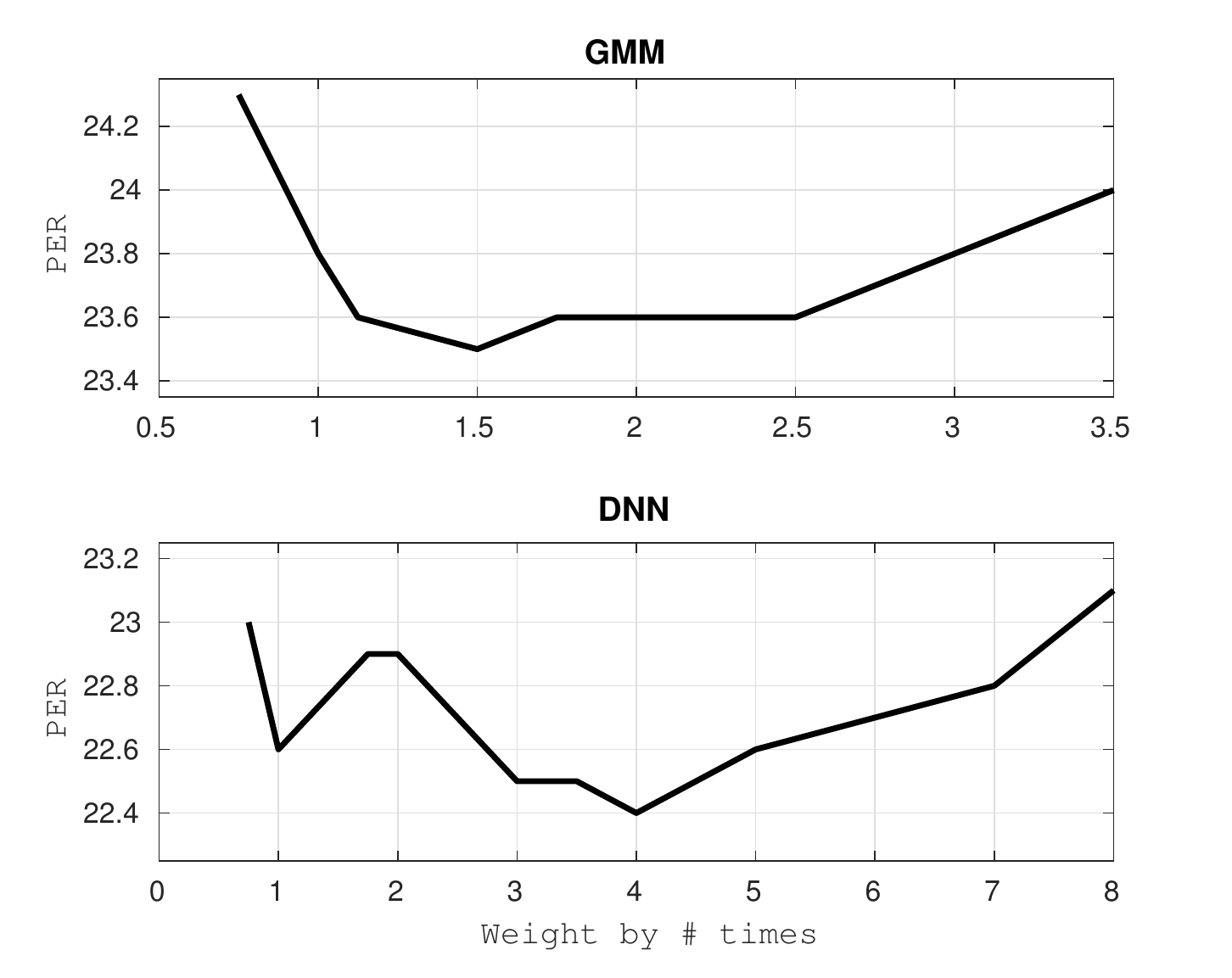}
      \caption{{\it Over-weighting landmark frames for GMM and DNN.}}
      \label{fig:reweight}
      \vspace{-2mm}
    \end{figure}
    %\vspace{-2mm}

\subsection{Methods of Replacement for Dropped Frames}
\label{sec:replacement}

  Figure~\ref{fig:dropFill} compares the performance of three \emph{methods of replacement}: {\tt Copy}, {\tt Fill\_0} and {\tt Fill\_const} when a {\tt Regular} frame dropping pattern is used. Results show that {\tt Fill\_0} and {\tt Fill\_const} suffer very similar PER increments as the percentage of frames dropped is increased, while {\tt Copy} shows a relatively smaller PER increment for drop rates of 40\% or 50\%. As for the comparison between acoustic models, DNN-based models outperform GMM-based at all drop rates. Notably, the {\tt Copy} approach synergizes well with DNN models, and is able to maintain low PER increments even up to 75\% drop rate; this finding is similar to findings reported in papers from~\citet{vanhoucke2013multiframe}.
  
  \begin{figure}[htbp]
      %\vspace{-2mm}
      \centering
      \includegraphics[width=\linewidth]{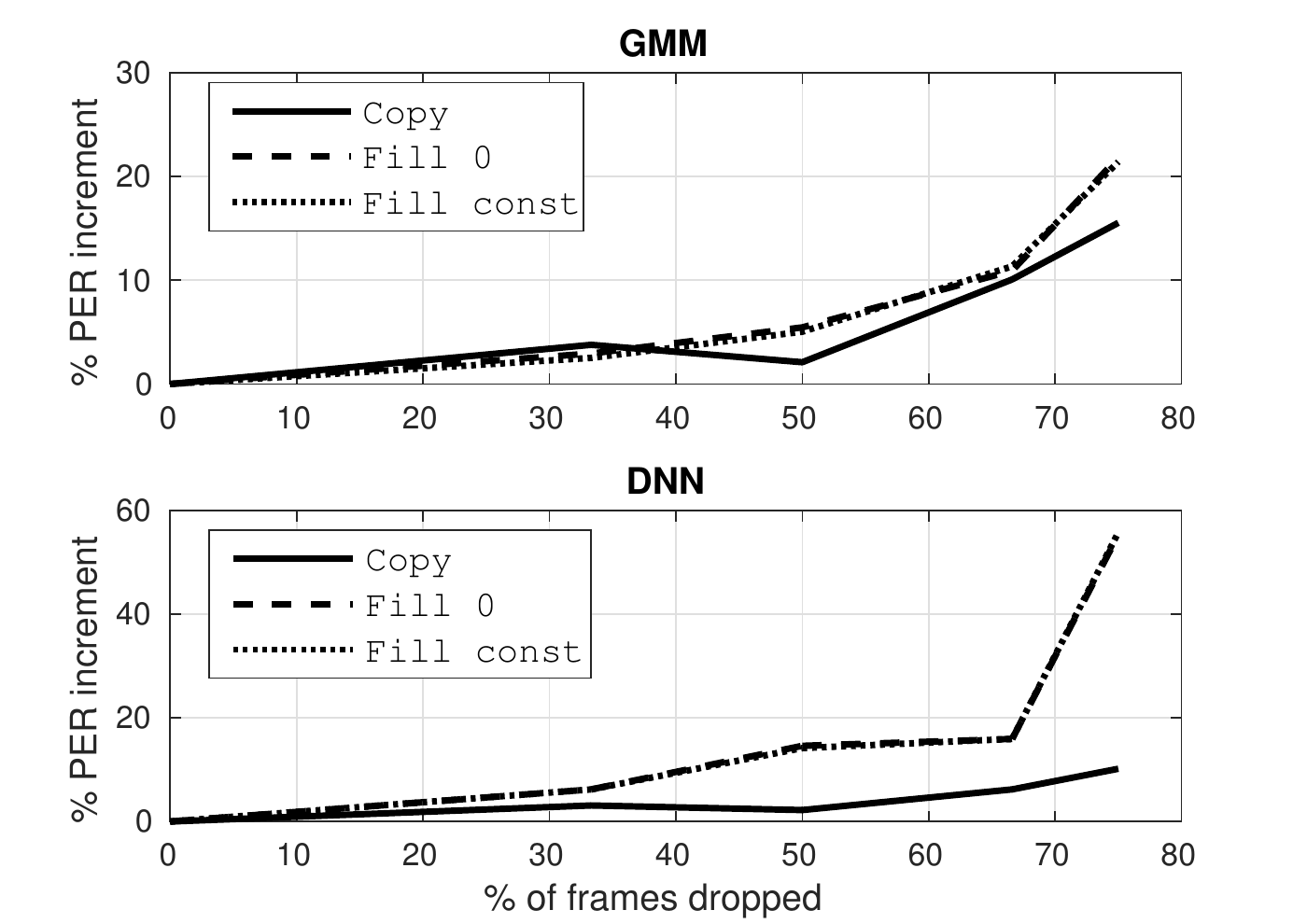}
      \caption{{\it Comparison of Different Methods of Frame Replacement ({\tt Copy}, {\tt Fill\_0} and {\tt Fill\_const}) assuming a {\tt Regular} pattern of frame replacement.} \label{fig:dropFill}}
      %\vspace{-2mm}
    \end{figure}

Figure~\ref{fig:dropType} compares the performance between two \emph{patterns of dropping frames} -- {\tt Regular}, {\tt Random}. In both of these the {\tt Copy} method for replacement was used. We also provide for comparison, the {\tt Regular} pattern, but using an {\tt Upsample} replacement method. This scheme uses a 17-tap anti-aliasing FIR filter. The method that offered the lowest phone error rate increment is obtained using a {\tt Regular} pattern with a {\tt Copy} replacement scheme. Results show that {\tt Regular-Copy} outperforms other methods by a large margin in terms of PER increment independent of which AM is used. 
  
    \begin{figure}[htbp]
      \centering
      \includegraphics[width=\linewidth]{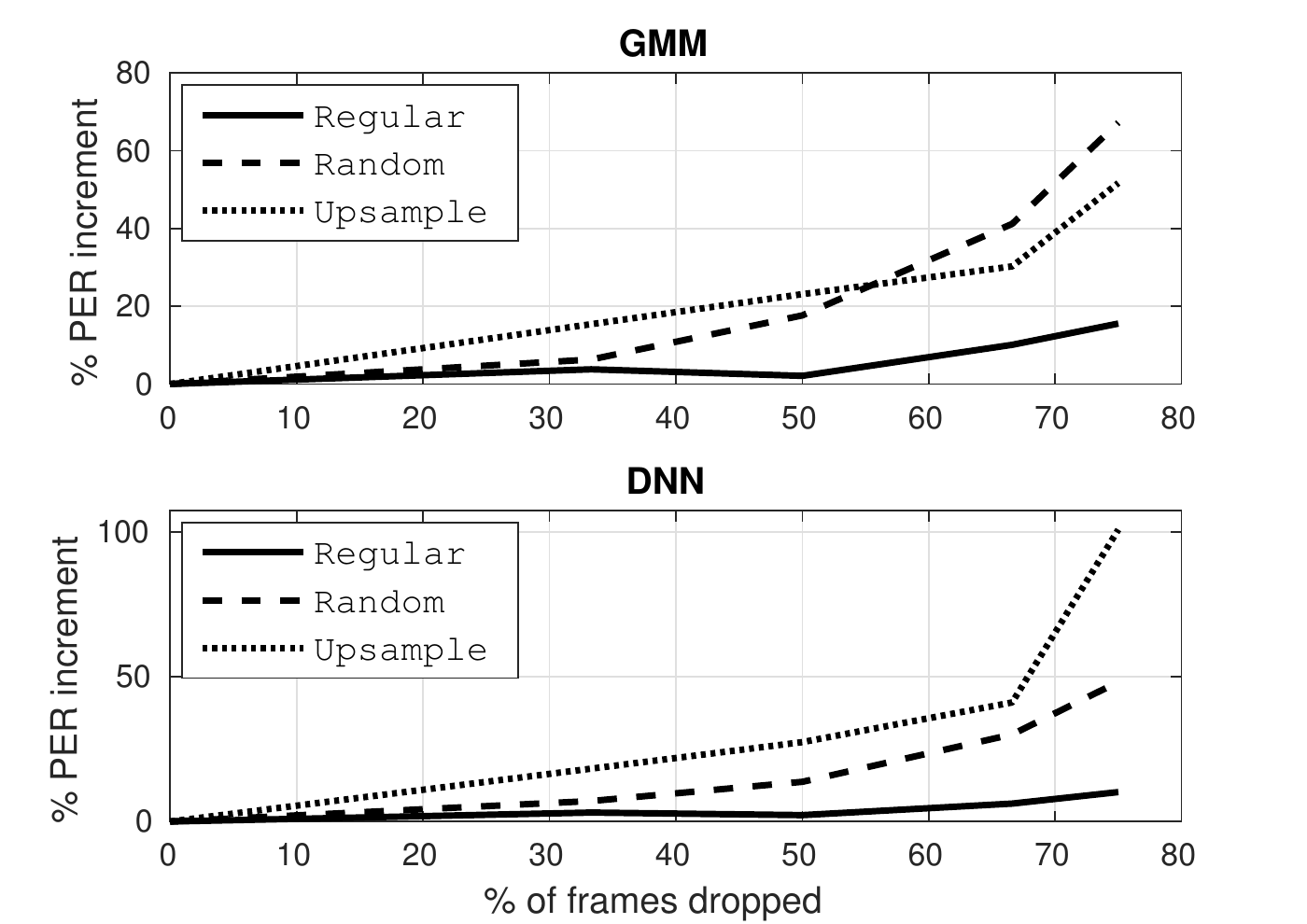}
      \caption{{\it Comparison of Different Patterns of Dropping Frames assuming {\tt Copy} ({\tt Regular} and {\tt Random}) and {\tt Interpolation through low-pass filtering} ({\tt Upsample}) method of replacement.}}
      \label{fig:dropType}
      %\vspace{-4mm}
    \end{figure}

  \subsection{Hypothesis 2: Dropping Frames with Regards to Landmarks}\label{subsec:dropping}
  
  At the beginning of this section, experiments that test hypothesis 2 directly are described. The focus is to subject the ASR decoding process to frames missing acoustic likelihood scores, and see how the decoding error rate changes accordingly. Obviously we are interested in using the presence vs.~absence of an acoustic landmark as a heuristic to choose the frames to keep or drop. To quantify the importance of the information kept vs.~the information discarded, dropping strategies ({\tt Landmark-keep} and {\tt Landmark-drop}) are compared to the non-landmark-based {\tt Random} strategy. Notice the {\tt Regular} strategy has been shown to be more effective than {\tt Random} (e.g., in Fig.~\ref{fig:dropType}), however, to make the PER result meaningful, the same number of frames should be dropped across different patterns being compared. When we keep only landmarks ({\tt Landmark-keep}) or drop only landmarks ({\tt Landmark-drop}), the percentage of frames dropped can not be precisely controlled by the system designer: it is possible to adjust the number of frames retained at each landmark (thus changing the drop rate), but it is not possible to change the number of landmarks in a given speech sample. Therefore, precisely adjusting the drop rate to meet a different pattern is not practical. Depending on the test set selected, the portion of frames containing landmarks ranges from $18.5\%$ to $20.5\%$. As opposed to {\tt Random}, {\tt Regular} does not give us the ability to select a drop rate that exactly matches the drop rate of the {\tt Landmark-drop} or {\tt Landmark-keep} strategies. Therefore, it is not covered in the first 2 experiments. However, in the 3rd experiment, we will compare a frame dropping strategy using landmark as heuristic against {\tt Regular} dropping. But that experiment will serve a slightly different purpose.
  
  As in the over-weighting experiment, two types of frame replacement are tested.  The {\tt Fill\_0} strategy is an exact implementation of hypothesis 2: when frames are dropped, they are replaced by the least informative possible replacement (a log probability of zero). Figure~\ref{fig:dropFill} showed, however, that the {\tt Copy} strategy is more effective in practice than the {\tt Fill\_0} strategy, therefore these two strategies are tested using a landmark-based frame drop pattern. Figure~\ref{fig:dropFill} showed that the~{\tt Fill\_const} strategy returns almost identical results to {\tt Fill\_0}, so it is not separately tested here.
  
  Experiment results are presented for both the TIMIT default test split, and for cross-validation (CV) using the whole corpus. The baseline implementation is as distributed with the Kaldi toolkit. Since no frames are dropped, it returns the lowest PER. However, likelihood scoring for the baseline AM will require more computation when compared to a system that drops frames. For CV we report the mean relative PER increment ($\Delta\mbox{PER}=100\times (\mbox{modified PER}-\mbox{baseline PER})/(\mbox{baseline PER})$), with its standard deviation in parentheses, across all folds of CV\@.  Every matching pair of frame-drop systems ({\tt Landmark-keep} versus {\tt Random}) is tested using a two-sample $t$-test~\citep{cressie1986use}, across folds of the CV, in order to determine whether the two PER increments differ. During the $t$-test, we assume PER numbers from different folds are samples of a random variable. The two-sample $t$-test intends to find out whether the random variables representing PER for different setups ({\tt Landmark-keep} versus {\tt Random}) have the same mean.
  
  \subsubsection{Keeping or Dropping the Landmark Frames}

Table~\ref{tab:table_1} illustrates the changes in PER increment that result from a {\tt Landmark-keep} strategy (score only landmark frames) versus a {\tt Random} frame-drop strategy set to retain the same percentage of frames. For each test set, we count the landmark frames separately and match the drop rate exactly between the {\tt Landmark-keep} and {\tt Random} strategy. In all cases, the {\tt Landmark-keep} strategy has a lower PER increment. A Wilcoxon test has been conducted on the default test set; the difference between all pairs but the DNN {\tt Fill0} pair is significant.

  \begin{table}[htbp]
    \centering
    % the first table in word doc
    \caption{{\it PER increments for scoring Landmark frames only compared to randomly dropping similar portion of frames (CV stands for cross validation; if the two increments differ, then the lower of the two is marked with either $\ast$ ($p<0.05$) or $\ast\ast$ ($p<0.001$).)}}
    %\vspace{2mm}
    \label{tab:table_1}
    \begin{tabular}{||l||c|c|c|c||c|c|c|c||}
    \hline
      Acoustic model 
      & \multicolumn{4}{c||}{GMM} & \multicolumn{4}{c||}{DNN}\\\hline
      Test regime
	& \multicolumn{2}{c|}{Default} 
    & \multicolumn{2}{c||}{CV Mean (Stdev)}
    & \multicolumn{2}{c|}{Default}
    & \multicolumn{2}{c||}{CV Mean (Stdev)} \\\hline
    Metric & PER & PER Inc & PER & PER Inc
    & PER & PER Inc & PER & PER Inc \\
    & (\%) & (\%) & (\%) & (\%) & (\%) & (\%) & (\%) & (\%) \\\hline
    Baseline & 23.8 & 0.0 & 22.8 & 0.0 
    & 22.7 & 0.0 & 20.8 & 0.0 \\ \hline
    \multicolumn{9}{||l||}{\tt Fill\_0} \\ \hline
    {\tt Landmark-keep} & 
36.1 & 51.7 & 33.4 & 46.5(1.34)** & 
49.6 & 118.5 & 49.7 & 139(10.3)*\\
      {\tt Random} 
      & 42.3 & 77.7 & 42.1 & 84.6 (8.35) 
      & 50.9 & 124.2 & 52.8 & 154 (14.8) \\
      \hline
    \multicolumn{9}{||l||}{\tt Copy} \\ \hline
    {\tt Landmark-keep} & 
35.2 & 47.7 & 32.3 & 41.5(1.08)** &
29.4 & 29.3 & 26.9 & 29.3(0.653)**\\
      {\tt Random} 
      & 44.0 & 84.9 & 44.1 & 93.5 (0.734) 
      & 38.4 & 69.3 & 37.6 & 80.9 (0.942) \\
      \hline
    \end{tabular}
  \end{table}

  For the next experiment we inverted the setup: instead of keeping only landmark frames, we drop only landmark frames (call this the {\tt Landmark-drop} strategy). Table~\ref{tab:table_2} compares the PER increment of a {\tt Landmark-drop} strategy to the increment suffered by a {\tt Random} frame drop strategy with the same percentage of lost frames. The {\tt Landmark-drop} strategy always return higher PER\@. However, only for the GMM setup {\tt Copy} did we obtain a significant $p$ value during cross validation. The $p$ values for other setups range from $0.13$ to $0.17$. Again, the Wilcoxon test has been conducted on the default test set, with the conclusion that only the GMM {\tt Copy} pair demonstrated significant difference.

  \begin{table}[htbp]
    \centering
    % the second table in table doc
    \caption{{\it PER increments for dropping Landmark frames during scoring compared to randomly dropping a similar portion of frames (CV stands for cross validation)}}
    %\vspace{2mm}
    \label{tab:table_2}
    \begin{tabular}{||l||c|c|c|c||c|c|c|c||}
    \hline
      Acoustic model
      & \multicolumn{4}{c||}{GMM} & \multicolumn{4}{c||}{DNN}\\\hline
      Test regime
	& \multicolumn{2}{c|}{Default} 
    & \multicolumn{2}{c||}{CV Mean (Stdev)}
    & \multicolumn{2}{c|}{Default}
    & \multicolumn{2}{c||}{CV Mean (Stdev)} \\\hline
    Metric & PER & PER Inc & PER & PER Inc
    & PER & PER Inc & PER & PER Inc\\ 
    & (\%)&(\%)&(\%)&(\%)&(\%)&(\%)&(\%)&(\%) \\\hline
    Baseline & 23.8 & 0.0 & 22.8 & 0.0 
    & 22.7 & 0.0 & 20.8 & 0.0 \\ \hline
    \multicolumn{9}{||l||}{\tt Fill\_0} \\ \hline
    {\tt Landmark-drop} & 
25.6 & 7.56 & 24.0 & 5.33(1.36) & 
24.2 & 6.61 & 23.1 & 11.1(1.58)\\
      {\tt Random} 
      & 24.1 & 1.26 & 23.4 & 2.68 (1.23) 
      & 23.6 & 3.96 & 22.4 & 7.53 (1.24) \\
      \hline
    \multicolumn{9}{||l||}{\tt Copy} \\ \hline
    {\tt Landmark-drop} & 
25.6 & 7.5 & 24.1 & 5.83(0.873)* &
24.3 & 7.1 & 22.1 & 6.44(0.836)\\
      {\tt Random} 
      & 24.6 & 3.3 & 23.1 & 1.14 (0.948) 
      & 23.6 & 4.0 & 21.6 & 3.85 (0.760) \\
      \hline
    \end{tabular}
  \end{table}

  The results in Table~\ref{tab:table_1} demonstrate that keeping landmark frames is better than keeping a random selection of frames at the same drop rate, in all but one of the tested comparison pairs.  The results in Table~\ref{tab:table_2} demonstrate that random selection tends to be better than
  selectively dropping the landmark frames, though the difference is only significant
  in one of the four comparison pairs.  These two findings support the hypothesis that frames containing landmarks are more important than others.  However, the PER increment in some setups are very large, indicating the ASR might no longer be functioning under stable conditions.
  
  \subsubsection{Using Landmark as a Heuristic to Achieve Computation Reduction}
  
  Methods in Table~\ref{tab:table_1} and~\ref{tab:table_2} compared the {\tt Landmark-keep}, {\tt Landmark-drop}, and {\tt Random} frame drop strategies.  Table~\ref{tab:landmarkAcc} illustrates PER increment (\%) for the {\tt Landmark-keep} and {\tt Regular} frame-dropping strategies. In this experiment, we are no longer directly testing Hypothesis 2. Instead, we are trying to achieve high frame dropping rate subject to low PER increment. As dropped frames need not be calculated during the acoustic model scoring procedure, a high dropping ratio can benefit the ASR by reducing computational load. The strategy leveraging landmark information is a hybrid strategy: on top of a standard {\tt Regular} strategy, it keeps all landmark frames and over-weights the likelihoods of these frames as in~\ref{sec:over}. For each acoustic model type (GMM vs.~DNN), three different percentage rates of frame dropping are exemplified.  In each case, we select a {\tt Regular} strategy with high dropping rate, modify it to keep the landmark frames, measure the percentage of frames dropped by the resulting strategy, then compare the result to a purely {\tt Regular} frame-drop strategy with a similar drop rate. The baseline {\tt Regular} strategies have three standard drop rates: 33.3\% (one out of three frames dropped, uniformly), 50\% (one out of two frames dropped), and 66.7\% (two out of three frames dropped). Table~\ref{tab:landmarkAcc} highlights results for one of the setups in bold, as that setup achieves a very good trade off between high dropping ratio and low PER increment.
  
  As we can see, for DNN acoustic models, the {\tt Landmark-keep} strategy results in lower error rate increment than a {\tt Regular} strategy dropping a similar number of frames. Wilcoxon tests demonstrated a statistically significant difference at all three drop rates. For GMM acoustic models, avoiding landmarks does not seem to return a lower error rate. In fact, the error rate is higher for 2 out of 3 different drop rates. The highlighted case in Table~\ref{tab:landmarkAcc} is intriguing because it the PER increment is so low, and this row will therefore serve as the basis for further experimentation in the next section. In this setup for DNN, over 50\% of the frames were dropped, but the PER only increased by 0.44\%. This result seems to support the hypothesis that landmark frames contain more information for ASR than other frames, but in Table~\ref{tab:landmarkAcc}, this row has the appearance of an anomaly, since the error increment is so small. In order to confirm that this specific data point is not a special case, we conducted additional experiments with very similar setups. The results for these additional experiments are presented in Table~\ref{tab:landmarkAcc2}.
  
  \begin{table}
    \centering
    \caption{{\it PER increments comparison between Landmark-keep and Regular drop strategies for GMM and DNN.}}
    \label{tab:landmarkAcc}
    %\vspace{2mm}
    %\begin{tabular}{|p{0.4cm}|p{1cm}|p{1.2cm}|p{1cm}|p{1.2cm}|p{1cm}|p{1.4cm}|p{1.4cm}|}
    \begin{tabular}{||c|c|r|r|r|r|r|r||}
      %\toprule
      \hline
      &\multirow{2}{*}{Copy} & \multicolumn{2}{c|}{Default} & \multicolumn{4}{c||}{Cross Validation} \\
      \cline{3-8}
      &&Drop Rate\% & PER Inc\%& Drop Rate\% & PER Inc\%& Inc STD\%& Inc pVal\\
      \hline
    \multirow{4}{*}{\rotatebox{90}{GMM}}  & Land & 41.0 & 1.26 & 44.4 & 1.84 & 0.0133 & 0.962 \\
      &Reg & 33.3 & 3.78 & 33.3 & 1.81 & 0.0119 & \\
      \cline{2-8}
      &Land & 54.2 & 2.94 & 54.1 & 2.86 & 0.0140 & 0.598 \\
      &Reg & 50 & 2.1 & 50 & 2.58 & 0.00780 & \\
      \cline{2-8}
      &Land & 64.3 & 12.1 & 65.0 & 8.10 & 0.0182 & 0.159 \\
      &Reg & 66.7 & 10.1 & 66.7 & 6.91 & 0.0181 & \\
      \hline
      \multirow{4}{*}{\rotatebox{90}{DNN}}  & Land & 41.0 & 0.44 & 44.4 & 1.84 & 0.0115 & 0.0011 \\
      &Reg & 33.3 & 3.98 & 33.3 & 4.20 & 0.0153 & \\
      \cline{2-8}
      & Land & \textbf{54.2} & \textbf{0.44} & \textbf{58.4} & \textbf{1.90} & \textbf{0.167} & \textbf{0.0029} \\
      &Reg & 50 & 2.21 & 50 & 4.12 & 0.0115 & \\
      \cline{2-8}
      &Land & 64.2 & 3.08 & 69.0 & 5.86 & 0.0121 & 0.0391 \\
      &Reg & 66.7 & 6.17 & 66.7 & 7.04 & 0.0160 & \\
      \hline
    \end{tabular}
  \end{table}
  
\begin{table}
\centering
\caption{{\it PER increments for Landmark-keeping strategy for DNN with dropping rate near 54.2\% and over-weighting factor near 4 times}}
\label{tab:landmarkAcc2}
\begin{tabular}{||l|l|p{1.5cm}|p{1.5cm}|p{1.5cm}||}
\hline
PER Inc\%                    &      & \multicolumn{3}{l||}{Over-weighting Factor} \\ \hline
                             &      & 3.5        & 4          & 4.5       \\ \hline
\multirow{3}{*}{Drop Rate\%} & 52.1 & 1.42       & 0.84       & 0.93      \\ \cline{2-5} 
                             & 54.2 & 0.88       & 0.44       & 0.88      \\ \cline{2-5} 
                             & 56.3 & 0.62       & 0.40       & 0.40      \\ \hline
\end{tabular}
\end{table}

  Additional results presented in Table~\ref{tab:landmarkAcc2} are obtained through applying an over-weighting factor close to 4, which is the optimal value found for DNNs in Figure~\ref{fig:reweight}. The first and third rows in this table randomly keep or drop a small number of non-landmark frames, in order to obtain drop rates of $52.1\%$ and $56.3\%$ respectively. Since the selection is random, multiple runs of the experiment result in 
different PER for the same drop rate, therefore we repeated each experiment 10 times and reported the mathematical mean. Since there is a level of randomness in these results, we do not intend to evaluate our hypotheses on these data; rather, the goal of Table~\ref{tab:landmarkAcc2} is merely to confirm that the highlighted case in Table~\ref{tab:landmarkAcc} is a relatively stable result of its parameter settings, and not an anomaly. Since good continuity can be observed across nearby settings, results in Table~\ref{tab:landmarkAcc2} lend support to the highlighted test case in Table~\ref{tab:landmarkAcc}.
  
  \section{Discussion}\label{sec:discussion}
  
Results in Section~\ref{sec:over} tend to support hypothesis 1. However, the tendency is not statistically significant. The tendency is consistent for the GMM-based system, for all over-weighting factors between 1.0 and 3.0. Similar tendencies appeared for over-weight factors between 3.0 and 5.0 for DNN-based system.

  Experiments in Section~\ref{sec:replacement} tested different non-landmark-based frame drop strategies, and different methods of frame replacement.  It was shown that, among the several strategies tested, the {\tt Regular-Copy} strategy obtains the smallest PER\@. There is an interesting synergy between the frame-drop strategy and the frame-replacement strategy, in that the PER of a 50\% {\tt Regular-Copy} system (one out of every two frames dropped) is even better than that of a 33\% {\tt Regular-Copy} system (one out of every three frames dropped). This result, although surprising, confirms a similar finding reported by~\citet{sak2014long}. We suspect that the reason may be relevant to the regularity of the 50\% drop rate. When we drop 1 frame out of every 2 frames, the effective time span of each remaining frame is 20ms, with the frame extracted at the center of the time span. Dropping 1 frame out of every 3 frames, on the other hand, results in an effective time span per frame of 15ms, but the alignment of each frame's signal window to its assigned time span alternates from frame to frame. 
  
  It is worth mentioning that our definition of acoustic landmarks differs from that of~\citet{lulich2010subglottal} -- specifically, Lulich claims that there is no landmark in the center of Vowel and Glide. Instead, formant-subglottal resonance crossing, which is known to sit between the boundaries of [-Back] and [+Back] vowels, contains a landmark. It is possible that an alternative definition of landmarks might lead to better results.

  We can also observe that GMM and DNN acoustic models tend to perform differently in the same setup. For example, for GMM, randomly dropping frames results in a higher PER than up-sampling; this is not the case for DNN models. Results also demonstrate that DNN models perform quite well when frames are missing. A PER increment of only 6\% occurs after throwing away $2/3$ of the frames. GMM models tend to do much worse, especially when the drop rate goes up. 
  
 All experiments on DNN tend to support the strategy to avoid dropping landmarks. However, the 2 test cases covered in Table~\ref{tab:table_2} lack statistical confidence. Scoring only the landmark frames (the {\tt Landmark-keep} strategy) out-performs both {\tt Random} and {\tt Regular} frame-drop-strategies. On the other hand, if landmark frames are dropped (the {\tt Landmark-drop} strategy), we obtain higher PER when compared to randomly scoring a similar number of frames. 

We find, at least for ASR with DNN acoustic models, that landmark frames contain information that is more useful to ASR than other frames.  In the most striking case,
the highlighted result in Table~\ref{tab:landmarkAcc} indicates that it is possible to drop more than 54\% of the frames but only observe a 0.44\%  increment in the PER compared to baseline (PER increases from 22.7 to 22.8). We conclude, for DNN-based ASR, that experiments support hypothesis 2 (with statistically significant differences in two out of the three comparisons).  In comparison, we failed to find support for hypothesis 2 in GMM-based ASR.
% * <boonpang.lim@gmail.com> 2017-08-05T04:56:44.687Z:
% 
% >  0.44\%
% This is absolute right, do we want to point out that the relative error is also small?
% 
% ^ <dihe2@illinois.edu> 2017-08-05T20:07:38.987Z.

%Different drop rates are necessary for GMM and DNN in order to achieve the best trade off. For DNN, around 54\% is ideal since there is almost no accuracy loss by chopping out more than half of the computation. Yet we have shown that the {\tt Landmark-keep} DNN still significantly outperforms {\tt Regular-Copy} up to drop rates around 66.7\%.  Overall, the {\tt Landmark-keep} strategy performs better on DNN than GMM.

 \subsection{How Landmarks Affect the Decoding Results} 

Having proven that the {\tt Landmark-keep} strategy is more effective than a {\tt Random}
or {\tt Regular} drop strategy, we proceeded to investigate the resulting changes in the rates of insertion, deletion and confusion among phones. We compared the normalized increment of each type of error, separately, when the confusion matrices of the baseline system are subtracted from the confusion matrices of the {\tt Landmark-keep} and {\tt Random} frame-drop systems. Fig.~\ref{fig:indel} compares the normalized error increment, of different types of errors, for the {\tt Landmark-keep} and {\tt Random} strategies. The numbers reported in the figure are normalized error increment. They are calculated using error increment divided by the occurrence of each kind of phone. We use this measure to reflect the increment ratio while avoiding having to deal with situations that could lead to division by zero.
    
\begin{figure}[!ht]
  \centering    
    \begin{subfigure}[b]{0.48\textwidth}
    \includegraphics[width=\textwidth]{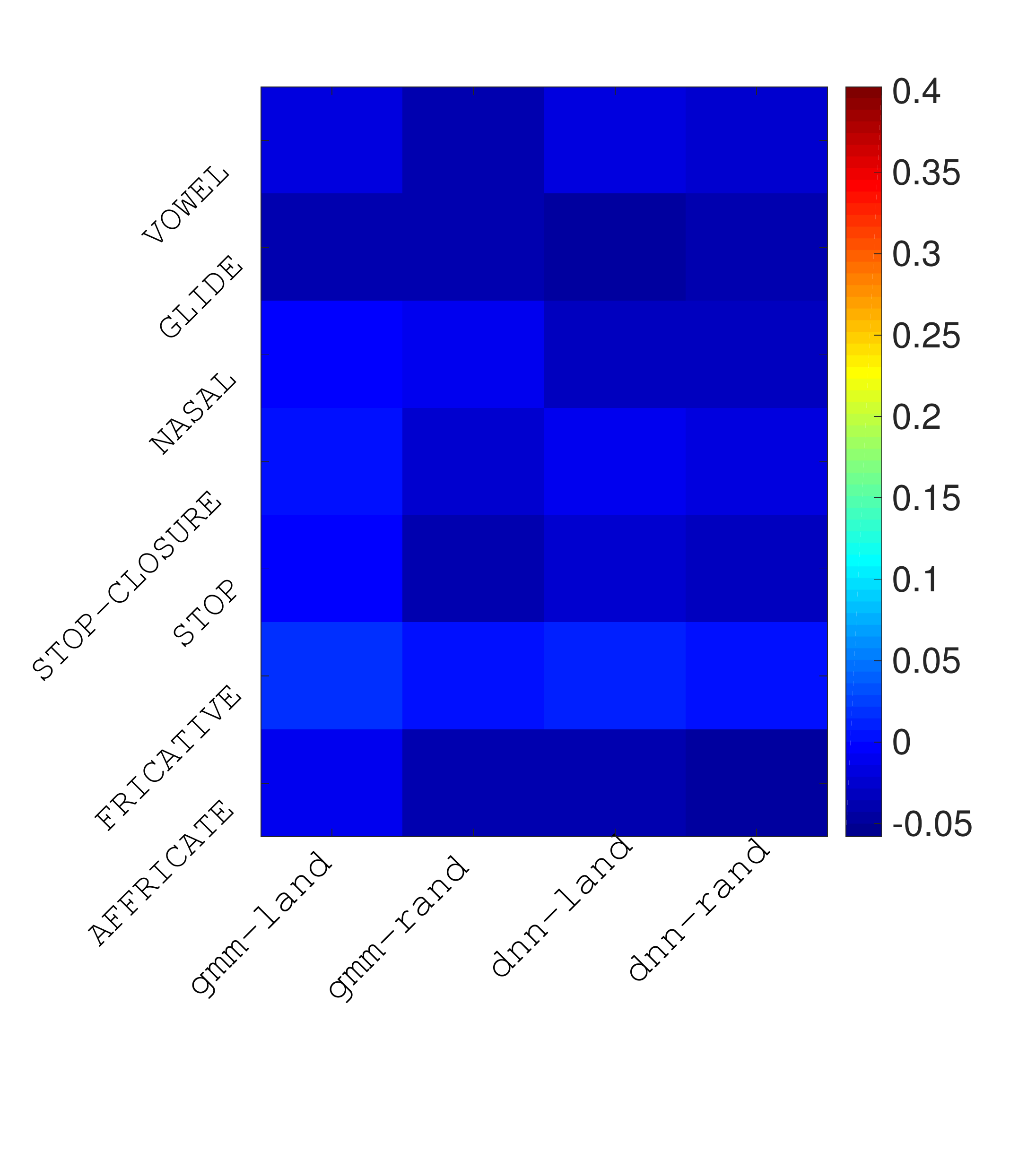}
    \caption{{\it Insertion errors}}
    \label{fig:ins7}
  \end{subfigure}
  ~
  \begin{subfigure}[b]{0.48\textwidth}
    \includegraphics[width=\textwidth]{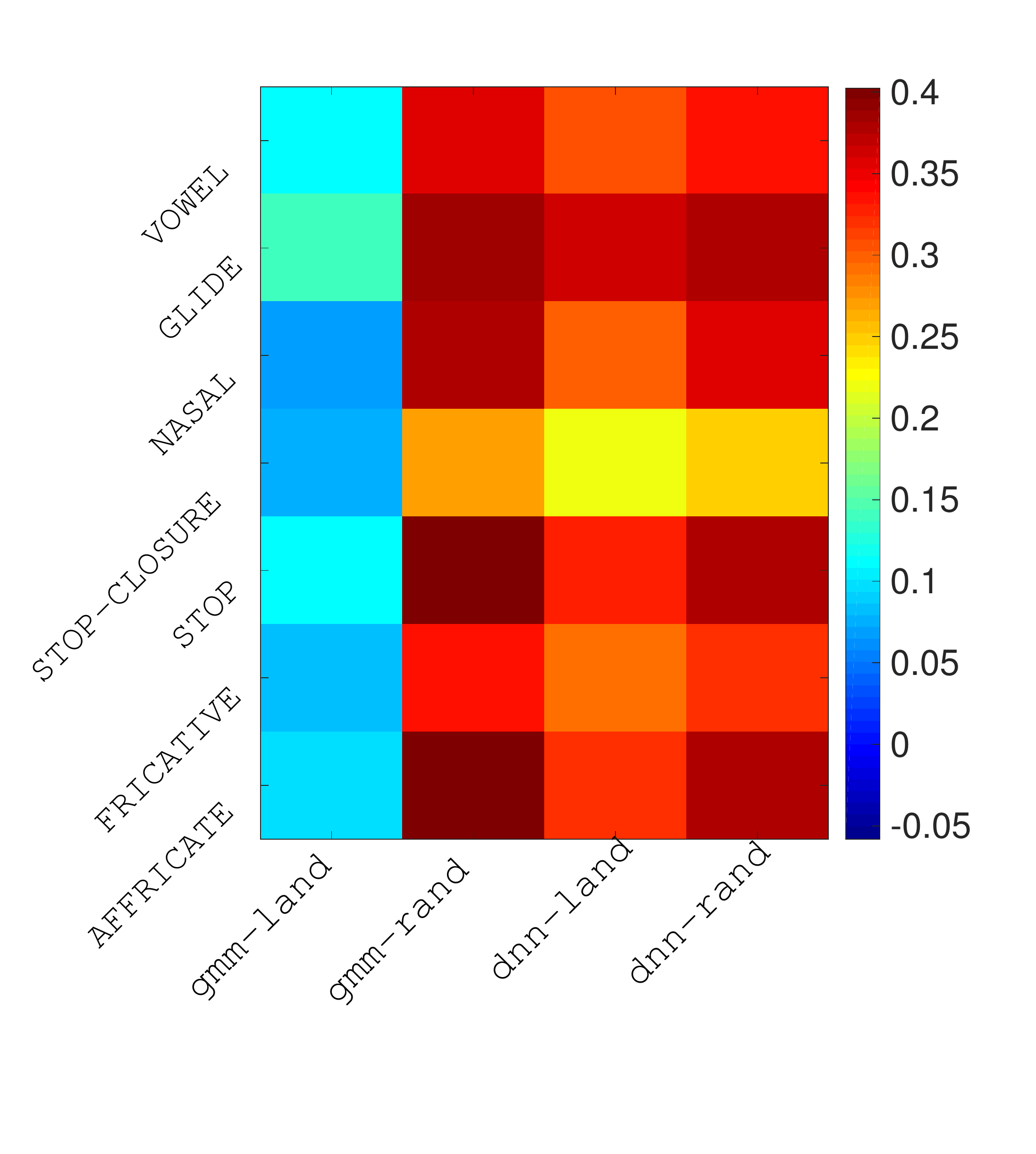}
    \caption{{\it Deletion errors}}
    \label{fig:del7}
  \end{subfigure}
  
  \caption{{\it The normalized error increment for a) insertion errors and b) deletion errors (y-axis represent different manners of articulators and x-axis represent different systems)}}
  \label{fig:indel}
\end{figure}

   Overall, dropping frames causes a minor reduction to the phone insertion rate, while the phone deletion rate significantly worsens. We suspect that after dropping frames, the decoder is less effective at capturing transitions between phones, resulting in correctly detected phones spanning over other phones. In Figure~\ref{fig:del7} we can see that the {\tt Landmark-keep} strategy is more effective than the {\tt Random} strategy, since it returns a lower deletion rate increment. We believe this is because the landmark contains sufficient acoustic information about each phone to force it to be recognized. However, we do not know why the GMM-{\tt{Landmark-keep}} strategy is less effective at preventing phone deletions than the DNN-{\tt{Landmark-keep}} strategy. A possible reason might be that more frames were stacked together in the splicing process for the DNN than for the GMM~\citep{Vesely13interspeech}
   %, --MH: I THINK THIS IS AN INCORRECT STATEMENT, OPPOSITE OF WHAT'S IN THE FIGURE? --- and as a result, the DNN may be more likely than the GMM to misclassify a landmark as one of the context phones rather than as the phone correctly associated with the landmark. 
   If we do consider providing landmarks as extra information to ASR, in order to reduce computation load for example, the difference between GMM and DNN models should be considered.

\section{Conclusions}\label{sec:conclusion}

Phones can be categorized using binary distinctive features, which can be extracted through acoustic cues anchored at acoustic landmarks in the speech utterance. In this work, we proved through experiments for DNN-based ASR systems operating on MFCC features, on the TIMIT corpus, using both the default and cross validation train-test splits, that frames containing landmarks are more informative than others. We proved that paying extra attention to these frames can potentially compensate for accuracy lost when dropping frames during Acoustic Model likelihood scoring. We leveraged the help of landmarks as a heuristic to guide frame dropping during speech recognition. In one setup, we dropped more than 54\% of the frames while adding only 0.44\% to the Phone Error Rate\@. This demonstrates the potential of landmarks for computational reduction for ASR systems with DNN acoustic models. We conclude that a DNN-based system is able to find a nearly-sufficient summary of the entire spectrogram in frames containing acoustic landmarks, in the sense that, if computational considerations require one to drop 50\% or more of all speech frames, one is better off keeping the landmark frames than keeping any other tested set of frames.  GMM-based experiments return mixed results, but results for the DNN are consistent and statistically significant: landmark frames contain more information about the phone string than frames without landmarks.

% --------------------------------------------------------------------------------------------------------------------------------

\end{space}
\vspace{2in}     %  included for better reading and an example of adding vertical space
%Two different styles included in the examples bibliography list.
\clearpage
\newpage

%-------------------------------------------------------------------------------------------------------------
% List of Figure Captions
%-------------------------------------------------------------------------------------------------------------
%%\begin{itemize}

% Figure1 -  no bullets
%% \item[]{Fig.~\ref{fig:stack}.  Stacking of Feature Frames Before the Scoring Process.} 

% Figure2
%% \item[]{Fig.~\ref{fig:reweight}. Overweighting landmark frames for GMM and DNN.} 
 
%% \item[]{Fig.~\ref{fig:dropFill}. Comparison of Different Methods of Frame Replacement ({\tt Copy}, {\tt Fill\_0} and {\tt Fill\_const}) assuming a {\tt Regular} pattern of frame replacement.}
 
%% \item[]{Fig.~\ref{fig:dropType}. Dropping frames in different manner.}
 
%% \item[]{Fig.~\ref{fig:indel}. Insertion and deletion (a) insertion, (b) deletion.}
  
%%\end{itemize}
  
%\end{space}

\end{document}